\documentclass[%
 reprint,
superscriptaddress,
nolongbibliography,
 amsmath,amssymb,
 aps,
floatfix,
]{revtex4-2}

\usepackage{graphicx}
\usepackage{dcolumn}
\usepackage{bm}
\usepackage{hyperref}

\usepackage{siunitx} 
\usepackage{physics} 
\usepackage{cleveref} 

\usepackage{multirow} 
\usepackage{booktabs} 
\usepackage{xcolor}
\usepackage[caption=false]{subfig}

\usepackage{xcolor}
\hypersetup{
	colorlinks=true,
	citecolor=blue,
	linkcolor=blue,
	urlcolor=blue,
}

\Crefname{equation}{Eq.}{Eqs.}
\Crefname{figure}{Fig.}{Fig.}
\newcommand{\nuclide}[2]{\ensuremath{^{#1}\mathrm{#2}}}
\newcommand{\minv}[1]{\ensuremath{{({m}^{-1})}_{#1}}}

\begin{document}

\title{Kinetic energy of fission fragments within a dynamical model}
\author{S.\,Takagi}
\affiliation{Graduate School of Science and Engineering, Kindai University, Higashi-Osaka, Osaka 577--8502, Japan}
\affiliation{Advanced Science Research Center, Japan Atomic Energy Agency, Tokai, Ibaraki 319--1195, Japan}
\author{Y.\,Aritomo}
\author{K.\,Nakajima}
\affiliation{Graduate School of Science and Engineering, Kindai University, Higashi-Osaka, Osaka 577--8502, Japan}
\author{K.\,Okada}
\author{K.\,Hirose}
\author{K.\,Nishio}
\affiliation{Advanced Science Research Center, Japan Atomic Energy Agency, Tokai, Ibaraki 319--1195, Japan}

\date{\today}
\begin{abstract}
	Kinetic energy of individual fission fragment for actinide nuclei is, for example, important for evaluating the prompt-neutron spectrum in the laboratory system.
	It is experimentally known that kinetic energy for each fragment is constant at about $\SI{100}{\MeV}$
	for light fragments and that for heavy fragments decreases linearly with mass number.
	Most of the theoretical studies carried out so far attempted to calculate the total kinetic energy of both fragments, i.e. sum of the energies of two fragments,
	but the kinetic energy of each fragment was not analyzed in detail as far as we recognize.
	We have calculated them in thermal-neutron induced fission of $\nuclide{239}{Pu}$ with a dynamical model using Langevin equations within a three-dimensional two-center parametrization.
	Also fission of $\nuclide{258}{Fm}$ was investigated.
	It is calculated from the Coulomb energy at the scission point and the pre-scission kinetic energy.
	It is found that the pre-scission kinetic energy has about 2-4\% contribution in the kinetic energy.
	The calculated results reproduce the trend of the experimental data.
\end{abstract}

\maketitle

\section{introduction}
There are many fission observables important for understanding the fission mechanism and atomic energy applications.
The data includes fission fragment mass distribution (FFMD) and total kinetic energy (TKE).
In the framework of the classical liquid drop model, the saddle point and fission valley are located at mass-symmetry, thus the FFMD is predicted to have a mass-symmetric shape.
By introducing the shell correction energy\ \cite{Strutinsky1967-ua, Strutinsky1968-qf}, the actinide nucleus splits into mass asymmetry, explaining the experimental data.
For a wide range of actinide nuclides, the average mass number of heavy fragments is known to be constant at about $A=140$\ \cite{Wagemans1991-kx, Vandenbosch1973-ev}.
Proton shell of $Z=54$ is found to have an important role in determining the degree of mass-asymmetry\ \cite{Schmidt2001-fz}.
This is theoretically interpreted \ \cite{Scamps2018-xe} as due to the onset of octupole (pear-shaped) deformation of the nascent heavy fragment \ \cite{Scamps2018-xe}.
The total kinetic energy mostly originates from the Coulomb energy at the scission point.

Here, we focus on the kinetic energy of individual fission fragment.
In actinide nuclides from thorium to californium,
the average kinetic energy of light fragments is known to have a nearly constant value from $\SI{99}{\MeV}$ to $\SI{107}{\MeV}$, depending on the nuclide,
whereas for heavy fragments it tends to decrease linearly with mass number\ \cite{Asghar1977-rd, Asghar1981-ek, Asghar1982-fn, Wagemans1981-zc, Wagemans1984-db, Wagemans1996-sx, Schmitt1965-us, Schmitt1966-iv, Nishio1995-wv, Caitucoli1981-hf, Caitucoli1983-kz, Meadows1969-wl, Baba1997-pj}.
In spite of the many attempts to calculate the TKE with different theories, individual kinetic energy has not been theoretically studied so far in detail.
Although Coulomb energy at the scission point is the dominant origin for kinetic energy, the pre-scission kinetic energy (PKE) may also contribute.
Among the various theoretical models, scission point model, for example, cannot introduce PKE because the calculation is performed to configure the shape at the scission point\ \cite{Wilkins1976-kj, Carjan2015-bv} without introducing the collective motion of nucleus.
On the other hand, the dynamical models such as the Langevin approach can evaluate the time evolution of nuclear shape, thus can define PKE for each fragment\ \cite{Aritomo2014-na, Ishizuka2017-ec}.
We calculated the individual kinetic energy by taking into account the PKE for thermal-neutron induced fission of $\nuclide{239}{Pu}$ in the framework of Langevin equations.
We also calculated the fission of $\nuclide{258}{Fm}$, known to have a sharp mass-symmetric FFMD with large TKE in spontaneous fission\ \cite{Hulet1986-cj, Hulet1989-kp, Hoffman1995-zr}.
The calculation results show that the kinetic energy of the individual fragments reproduces the experimental trend as well as FFMD and TKE distribution.

\section{model}
We adopted the three-dimensional Langevin equations to calculate the time evolution of nuclear shape.
The nuclear shape is defined by the two center parametrization\ \cite{Maruhn1972-rv, Sato1978-jy}, which has three deformation parameters, $z_0$, $\delta$, and $\alpha$.
The symbol $z_0$ is the distance between the center of two potentials, $\delta$ denotes the deformation parameter of each fragment ($\delta = \delta_1 = \delta_2$), and $\alpha$ is the mass asymmetry of fragments.
The parameter $\delta$ is defined as $\delta = \frac{3(a - b)}{2a + b}$ using the half length of the ellipse axes $a$ (symmetry-axis direction) and $b$ (radius direction).
The mass asymmetry $\alpha$ is defined as $\alpha = \frac{A_1 - A_2}{A_1 + A_2}$, represented by the fragment mass $A_1$ and $A_2$.
In addition, we use scaling to reduce computation time and introduce the coordinate $z$ defined as $z = \frac{z_0}{R_{\mathrm{CN}}B}$,
where $R_{\mathrm{CN}}$ indicates the radius of the spherical compound nucleus and $B$ is defined as $B = \frac{3 + \delta}{3 - 2\delta}$.
These three collective coordinates are abbreviated as $q$, with $q = \{z, \delta, \alpha \}$.
We use the neck parameter $\varepsilon = 0.35$ for $^{240}\mathrm{Pu}$ and $\varepsilon = 0.66$ for $^{258}\mathrm{Fm}$\ \cite{Miyamoto2019-gd}, determined to reproduce the experimental FFMDs.

The potential energy is a sum of the liquid drop (LDM) and shell correction (SH) parts.
\begin{equation} \label{eq:potential}
	\begin{aligned}
		V(q,T)                 & = V_{\mathrm{LDM}}(q) + V_{\mathrm{SH}}(q,T),                 \\
		V_{\mathrm{LDM}}(q)    & = E_\mathrm{S}(q)+E_\mathrm{C}(q),                            \\
		V_{\mathrm{SH}}(q,T)   & = E_{\mathrm{SH}}^{0}(q)\Phi(T),                              \\
		E_{\mathrm{SH}}^{0}(q) & = \Delta{E_{\mathrm{shell}}}(q) + \Delta{E_\mathrm{pair}}(q), \\
		\Phi(T)                & = \exp(-\frac{aT^2}{E_d}).
	\end{aligned}
\end{equation}
In \Cref{eq:potential}, the $V_{\mathrm{LDM}}$ is the potential from the finite-range liquid drop model\ \cite{Krappe1979-nk},
given as a sum of the surface energy $E_\mathrm{S}(q)$ and the Coulomb energy $E_\mathrm{C}(q)$.
The microscopic potential $V_{\mathrm{SH}}$ at $T=0$ is calculated as a sum of the shell correction energy $\Delta{E_{\mathrm{shell}}}(q)$, evaluated by the Strutinsky method\ \cite{Strutinsky1967-ua, Strutinsky1968-qf},
and the pairing correction energy $\Delta{E_\mathrm{pair}}(q)$\ \cite{Nilsson1969-gc, Brack1972-qr}.
$T$ is the temperature of the nucleus calculated from the intrinsic energy of the composite system.
The temperature dependence factor $\Phi(T)$ is discussed in Ref.\ \cite{Aritomo2004-tj}, calculated using shell damping energy ($E_\mathrm{d} =$ $\SI{20}{\MeV}$ ) given by Ignatyuk \textit{et al.}\ \cite{Ignatyuk1975-iw}.

The Langevin equation\ \cite{Aritomo2004-tj} is given as
\begin{equation} \label{eq:langevin}
	\begin{aligned}
		\dv{q_i}{t} & = {\qty(m^{-1})}_{ij}p_{j},                                    \\
		\dv{p_i}{t} & = -\pdv{V}{q_{i}} - \frac{1}{2}\pdv{q_{i}} \minv{jk}p_{j}p_{k} \\
		            & \phantom{{}={}} - \gamma_{ij}\minv{jk}p_{k} + g_{ij}R_{j}(t),
	\end{aligned}
\end{equation}
where $p_i = m_{ij}\dv{q_j}{t}$ denotes the momentum conjugated to $q_i$.
In \Cref{eq:langevin}, $m_{ij}$ and $\gamma_{ij}$ are the shape-dependent collective inertia and friction tensors.
We adopted the hydrodynamical inertia tensor $m_{ij}$ in the Werner-Wheeler approximation\ \cite{Davies1976-yi}.
The one-body friction tensor $\gamma_{ij}$ is calculated in the wall-and-window formula\ \cite{Randrup1984-pc, Sierk1980-im}.
The normalized random force tensor $R_{i}(t)$ is assumed to be white noise, $\ev{R_{i}(t)} = 0$ and $\ev{R_{i}(t_{1})}\ev{R_{j}(t_{2})} = 2\delta_{ij}\delta(t_{1} - t_{2})$.
The strength of random force $g_{ij}$ is related to the friction tensor by the classical Einstein relation\ \cite{Hofmann1998-tv},
\begin{equation} \label{eq:effective_temperature}
	\begin{aligned}
		\sum_{k}g_{ik}g_{jk} & = \gamma_{ij}T^{*},                                    \\
		T^{*}                & = \frac{\hbar\omega}{2} \coth{\frac{\hbar\omega}{2T}}.
	\end{aligned}
\end{equation}
Here, $T^{*}$ is the effective temperature\ \cite{Randrup1979-sc,Aguiar1989-rz}.
The parameter $\omega$ is the local frequency of collective motion\ \cite{Hofmann1998-tv}.
The minimum of $T^*$ is given by $\hbar\omega / 2$, which corresponds to the zero-point energy of the oscillators that form the heat bath.
We estimated the zero-point energy as 1.4 MeV to be consistent with experimental data.
The temperature $T$ is related to the intrinsic energy $E_{\mathrm{int}}$ of the composite system as $E_{\mathrm{int}} = aT^2$, where $a$ is the level density parameter.
$E_{\mathrm{int}}$ is calculated at each step of a trajectory calculation,
\begin{equation}
	E_{\mathrm{int}} = E^* - \frac{1}{2}\minv{ij}p_{i}p_{j} - V(q, T=0). \label{eq:intrinsic}
\end{equation}
The excitation energy of the compound nucleus $E^*$ is given as \SI{6.53}{\MeV} for $\nuclide{240}{Pu}$, corresponding to thermal-neutron induced fission of $\nuclide{239}{Pu}$, and \SI{7.0}{\MeV} for $\nuclide{258}{Fm}$.

\begin{figure}[ht]
	\centering
	\includegraphics[width=1.0\linewidth]{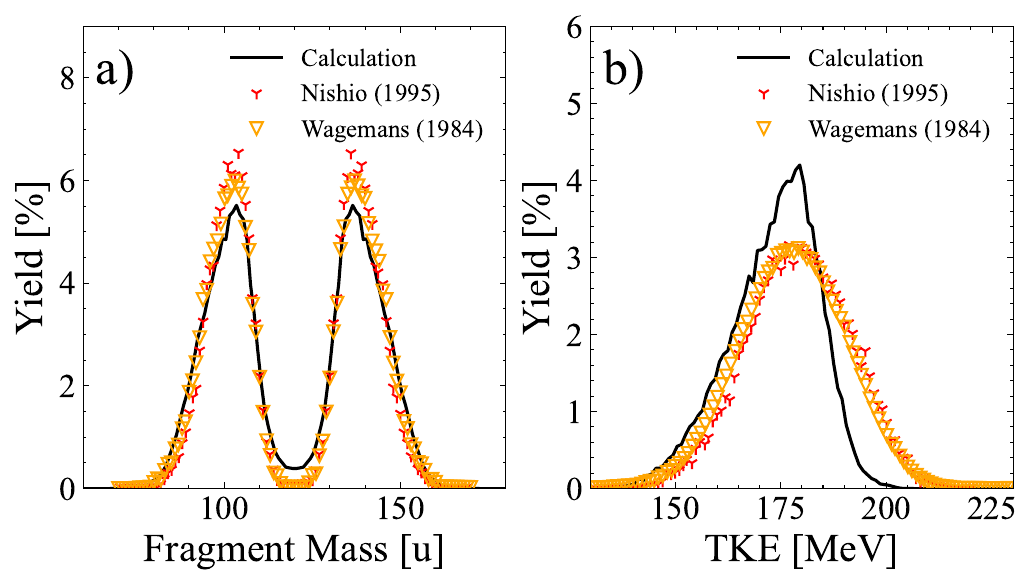}
	\caption{
		(a) FFMD and (b) TKE distribution for $^{240}\mathrm{Pu}$ in $E^{*}=\SI{6.53}{\MeV}$.
		The calculated results are shown by black lines.
		The experimental data of $^{239}\mathrm{Pu}(n_{\mathrm{th}}, f)$\ \cite{Nishio1995-wv, Wagemans1984-db} are shown by red and orange symbols ($\textsf{Y}, \triangledown$).
	}\label{fig:ffmd_tke}
\end{figure}

The total kinetic energy of the fission fragments can be expressed as a sum of the total Coulomb energy at the scission point $V_{\mathrm{Coul}}$ (TCE)
and pre-scission total kinetic energy of both fragments $E_{\mathrm{pre}}$ (PKE),
\begin{equation}
	\begin{aligned}
		\mathrm{TKE} & = V_{\mathrm{Coul}} + E_{\mathrm{pre}},                                                                \\
		             & = \frac{Z_{1}Z_{2}e^{2}}{D} + \frac{1}{2}\mu {\qty(\dv{z_{G2}}{t} - \dv{z_{G1}}{t})}^2 \label{eq:tke}.
	\end{aligned}
\end{equation}
$Z_1$ and $Z_2$ are the nuclear charge of fission fragments.
$D$ is the center-of-mass distance between the fragments at the scission point, defined by the shape that the neck-radius becomes zero.
$\mu$ is the reduced mass of the fragments, and
$z_{G1}$ and $z_{G2}$ are the center-of-mass coordinates of the light and heavy fragments, respectively.

Finally, the kinetic energy of each fragment (FKE) is calculated as follows using the momentum conservation law.
\begin{equation} \label{eq:fke}
	\begin{aligned}
		\mathrm{FKE}_{L} & = \frac{A_{H}}{A_{L}+A_{H}}\mathrm{TKE}, \\
		\mathrm{FKE}_{H} & = \frac{A_{L}}{A_{L}+A_{H}}\mathrm{TKE},
	\end{aligned}
\end{equation}
where $A_{L}$ and $A_{H}$ are the mass of light and heavy fragments, respectively.
Although our three-dimensional Langevin model cannot individually treat deformation of each fragment, reasonably predicted FFMD and TKE distributions shown below support the concept of calculating individual kinetic energy as we only introduce momentum conservation law.

\section{Result and Discussion}
\subsection{Fission properties of $\nuclide{240}{Pu}$}

\begin{figure}[htb]
	\centering
	\includegraphics[width=1.0\linewidth]{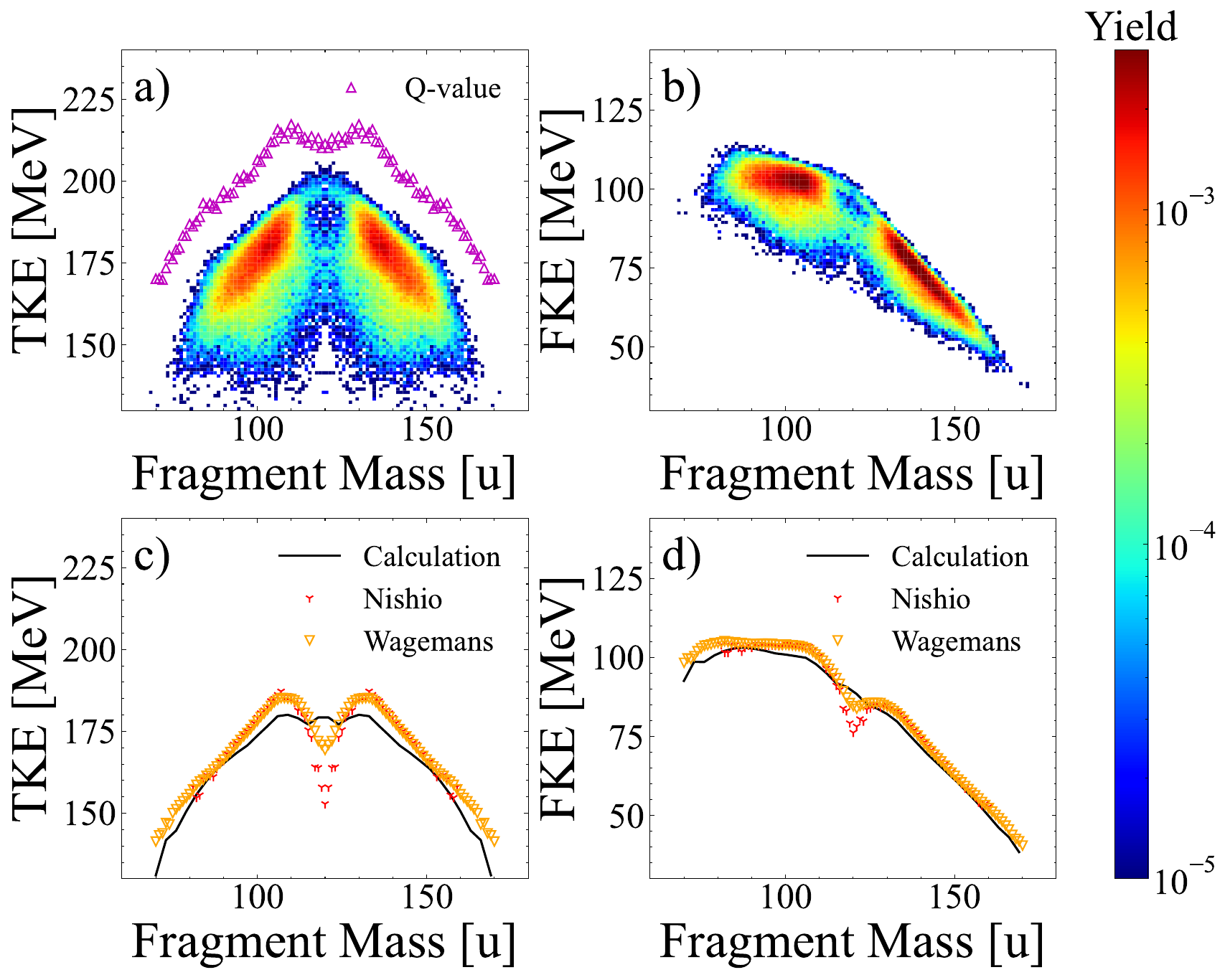}
	\caption{
		Calculated fission-fragment distributions on (a) mass-TKE and (b) mass-FKE for \nuclide{240}{Pu} at excitation energy of \SI{6.53}{\MeV}.
		A purple triangle in (a) is the Q-value.
		The calculated average TKE and average FKE are shown by solid lines in (c) and (d), respectively.
		The red and orange symbols ($\textsf{Y}, \triangledown$) are the experimental data of $\nuclide{239}{Pu}(n_{\mathrm{th}}, f)$\ \cite{Nishio1995-wv, Wagemans1996-sx}.
	}\label{fig:masstke_massfke}
\end{figure}

Figure\ \ref{fig:ffmd_tke} shows the calculated (a) FFMD and (b) TKE distribution, $Y(E_k)$, for fission of $\nuclide{240}{Pu}$ at the excitation energy $E^* = \SI{6.53}{\MeV}$, which are compared with the experimental data of $\nuclide{239}{Pu}(n_{th}, f)$\ \cite{Nishio1995-wv,Wagemans1996-sx}.
For FFMD, our calculation shows some deviation by showing larger yield in symmetric fission and lower yield at the mass-asymmetry giving the largest fraction.
In general, however, the calculation reproduces the overall shape of FFMD.
For the $Y(E_k)$ distribution, our calculation shows more narrow width than the measurement with larger yield around $E_k = \SI{180}{\MeV}$ and smaller yield for $E_k > \SI{185}{\MeV}$.
Still this discrepancy does not influence the main message given below.

Figure\ \ref{fig:masstke_massfke}(a) shows the calculated fission events on the fragment mass and TKE.
The fission Q-value, determined from the evaluated masses\ \cite{Moller2016-cy}, added by the excitation energy of the compound nucleus, is also shown.
The events distribute around $\SI{185}{\MeV}$ at heavy- and light-fragment masses of 135 and \SI{105}{\atomicmassunit} with a small yield at the symmetric fission.
Panel (b) is the distribution on the mass and FKE plane.
The most probable FKE is centered around $\SI{105}{\MeV}$ for the light fragments, but it linearly decreases with the mass number for the heavy fragments.
In Fig.\ \ref{fig:masstke_massfke} we show (c) average TKE and (d) average FKE as a function of fragment mass.
The results are compared with the experimental data\ \cite{Nishio1995-wv, Wagemans1996-sx}.
The experimentally observed average TKE in symmetric fission having a significant local minimum is overestimated by the calculation.
Also, the calculated average TKE at $A_{H} = 132$ predicts smaller value than the measurement.
However, the region $A_{H} = 140$ having the most dominant yield is explained in the calculation.
For the average FKE, the experimental data are reproduced in our calculation.
Especially, the constant values for the light fragments and linearly decreasing trend for heavy fragments are well demonstrated.

\begin{figure}[htb]
	\centering
	\includegraphics[width=1.0\linewidth]{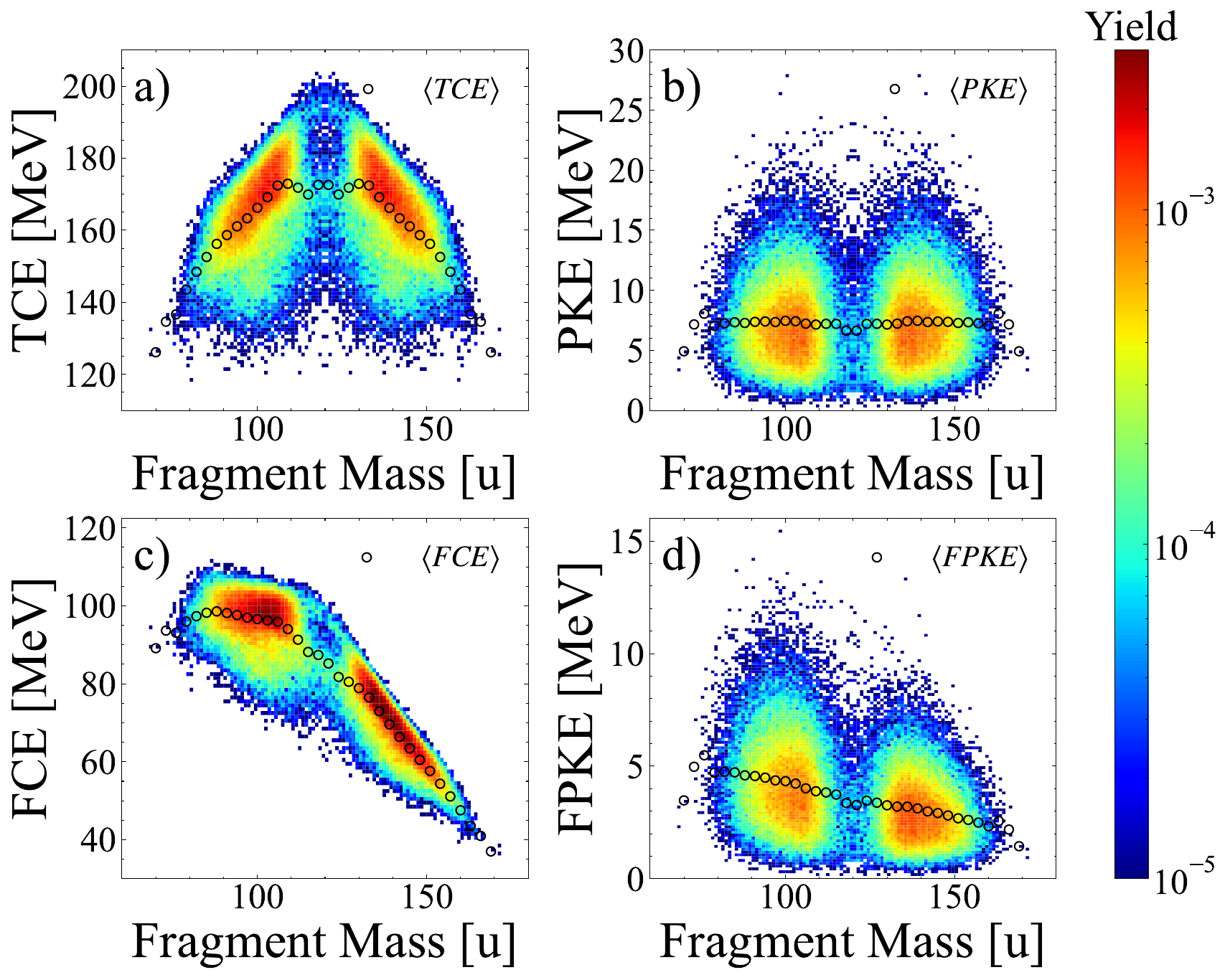}
	\caption{
		Fission fragment (a) mass-TCE, (b) mass-PKE, (c) mass-FCE, and (d) mass-FPKE distributions for $^{240}\mathrm{Pu}$ at excitation energy of \SI{6.53}{MeV}.
		The averages of each distribution are shown as open circles.
	}\label{fig:masstce_masspke}
\end{figure}

Figure\ \ref{fig:masstce_masspke} shows the (a) mass-TCE and (b) mass-PKE plots.
The black open circles indicate the average value for each fragment.
The mass-TCE distribution shows a similar distribution to the mass-TKE.
The mass-TCE distribution is about $\SI{7.0}{\MeV}$ shifted to lower values than mass-TKE in \Cref{fig:masstke_massfke}(a).
This difference comes from the PKE of both fragments shown in \Cref{fig:masstce_masspke}(b), whose average values are almost independent of the mass asymmetry.
Overall, the PKE has a 4.2\% contribution to the TKE value.
Figure\ \ref{fig:masstce_masspke} shows the (c) Coulomb part for each-fragment kinetic energy (FCE) and the (d) pre-scission kinetic energy for each fragment (FPKE).
The FCE for light fragments is distributed in the restricted region between $\SI{85}{\MeV}$ and $\SI{107}{\MeV}$, whereas the heavy fragment change its value widely from $\SI{85}{\MeV}$ at $A_H=\SI{125}{\atomicmassunit}$ to $\SI{45}{\MeV}$ at $\SI{165}{\atomicmassunit}$.
The mean value of FPKE decreases linearly with fragment mass from $\SI{5}{\MeV}$ down to $\SI{2}{\MeV}$.

\setlength{\tabcolsep}{8pt}
\renewcommand{\arraystretch}{1.7}
\begin{table*}[hbtp]
	\caption{
		Comparison of the calculation results for $^{240}\mathrm{Pu}$ ($E^* = \SI{6.53}{\MeV}$) with the experimental data for $^{239}\mathrm{Pu}(n_{\mathrm{th}},f)$.
		$\ev{A}$ indicates the average mass number of light/heavy fission fragments.
		$\ev{\mathrm{TKE}}$, $\ev{\mathrm{TCE}}$, and $\ev{\mathrm{PKE}}$ are average values of 
        total kinetic energy, total Coulomb energy, and pre-scission kinetic energy.
		$\ev{\mathrm{FKE}}$, $\ev{\mathrm{FCE}}$, and $\ev{\mathrm{FPKE}}$ denote average fragment kinetic energy, fragment Coulomb energy, and fragment pre-scission kinetic energy, respectively.
	}\label{table:cal_res}
	\centering
	\begin{tabular}{llccccccc}
		\toprule
		                                              & Fragment & \multicolumn{1}{c}{$\ev{A}$} & \multicolumn{1}{c}{$\ev{\mathrm{TKE}}$} & \multicolumn{1}{c}{$\ev{\mathrm{FKE}}$} & \multicolumn{1}{l}{$\ev{\mathrm{TCE}}$} & \multicolumn{1}{l}{$\ev{\mathrm{FCE}}$} & \multicolumn{1}{l}{$\ev{\mathrm{PKE}}$} & \multicolumn{1}{l}{$\ev{\mathrm{FPKE}}$} \\
		                                              &          & [u]                          & [MeV]                                   & [MeV]                                   & [MeV]                                   & [MeV]                                   & [MeV]                                   & [MeV]                                    \\
		\midrule
		\multirow{2}{*}{Calculation}                  & Light    & 100.5                        & \multirow{2}{*}{173.2}                  & 100.4                                   & \multirow{2}{*}{165.9}                  & 96.2                                    & \multirow{2}{*}{7.3}                    & 4.3                                      \\
		                                              & Heavy    & 139.5                        &                                         & 72.8                                    &                                         & 69.7                                    &                                         & 3.0                                      \\
		\multirow{2}{*}{Exp.\ \cite{Nishio1995-wv}}   & Light    & $100.9 \pm 0.7$              & \multirow{2}{*}{$178.9 \pm 1.2$}        & $103.5 \pm 1.0$                         &                                         &                                         &                                         &                                          \\
		                                              & Heavy    & $139.1 \pm 1.0$              &                                         & $75.4 \pm 0.7$                          &                                         &                                         &                                         &                                          \\
		\multirow{2}{*}{Exp.\ \cite{Wagemans1984-db}} & Light    & $100.30 \pm 0.01$            & \multirow{2}{*}{$177.65 \pm 0.1$}       & $103.29 \pm 0.01$                       &                                         &                                         &                                         &                                          \\
		                                              & Heavy    & $139.7  \pm 0.01$            &                                         & $74.36 \pm 0.01$                        &                                         &                                         &                                         &                                          \\
		\bottomrule
	\end{tabular}
\end{table*}

Table\ \ref{table:cal_res} summarizes the calculated results of the average values of the mass number, TKE, and FKE for the light and heavy fragments of $\nuclide{240}{Pu}$($E^* = \SI{6.53}{\MeV}$).
The results are compared to the experimental data of $\nuclide{239}{Pu}(n_{th}, f)$.
The calculated average masses of the heavy and light fragments well reproduced the data.
The calculated TKE and FKE are only a few percent smaller than the experimental data.
Table\ \ref{table:cal_res} also shows the calculated results of the mean values of the TCE, FCE, PKE, and FPKE.

\subsection{Fission properties of \nuclide{258}{Fm}}
\begin{figure}[htb]
	\centering
	\includegraphics[width=1.0\linewidth]{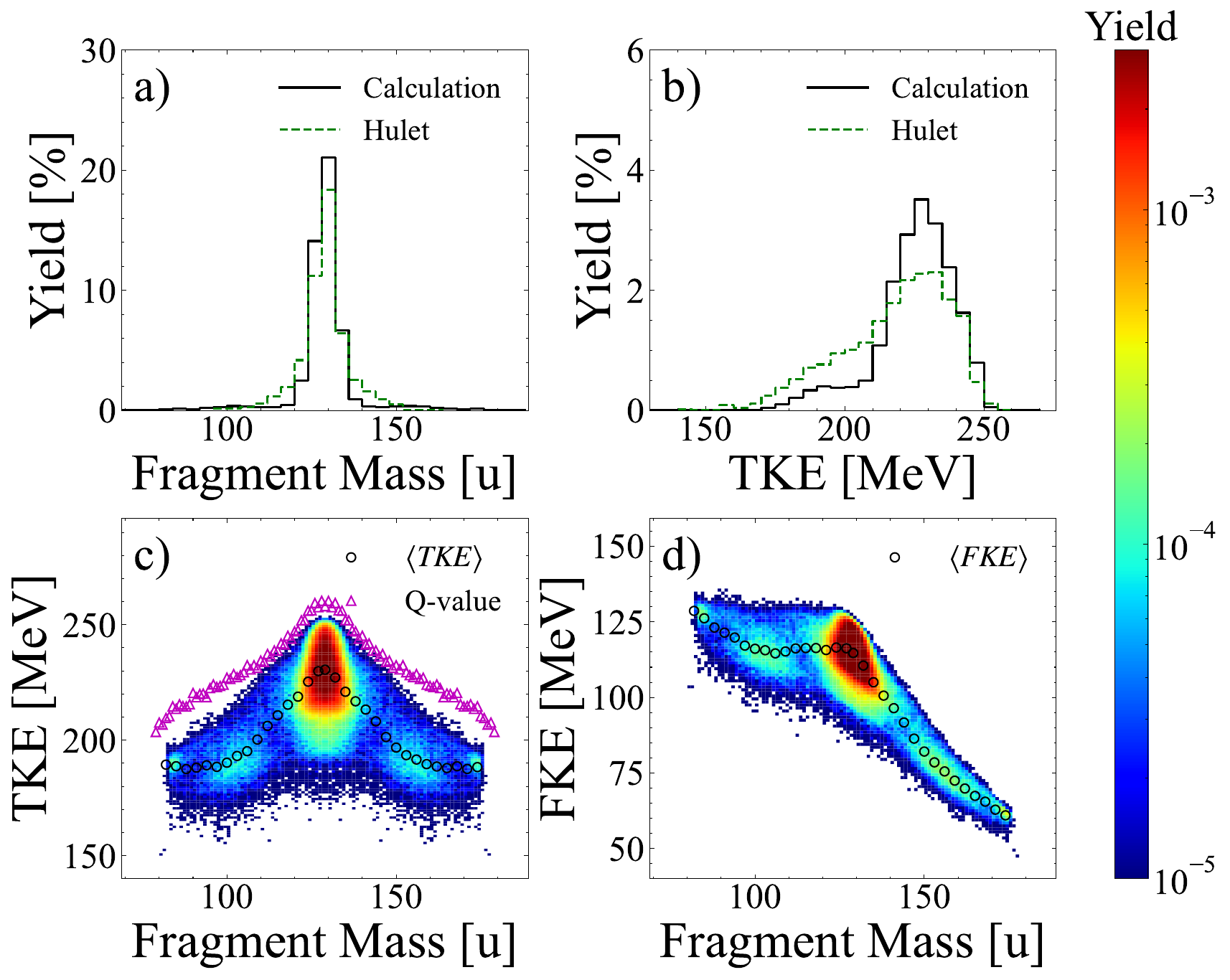}
	\caption{
		Calculated distributions of (a) FFMD, (b) TKE, (c) mass-TKE, and (d) mass-FKE for $\nuclide{258}{Fm}$ from $E^{*}=\SI{7}{\MeV}$.
		The FFMD and TKE of the present calculation are shown by black histograms and the experimental data of $^{258}\mathrm{Fm}(\text{sf})$\ \cite{Hulet1989-kp} is shown by green histograms.
		A purple triangle in (c) is the Q-value with the addition of the excitation energy. 
		In the panels of (c) and (d), the average value for each fragment is shown by black open circles.
	}\label{fig:masstke_massfke_fm258}
\end{figure}

In the same approach, we attempted the calculation of $\nuclide{258}{Fm}$ ($E^* = \SI{7.0}{\MeV}$).
Our model based on the semi-classical approach allows to treat only above-barrier fission.
Still comparison with measured spontaneous-fission data $\nuclide{258}{Fm}$ would be interesting to see how individual kinetic energy of fragment behaves as a function of fragment mass in our model.
Figure\ \ref{fig:masstke_massfke_fm258} shows the calculated (a) FFMD, and (b) TKE distribution, $Y(E_k)$, in comparison with the experimental data of\ \cite{Hulet1989-kp}.
The black histograms are the results of the present calculations.
The calculated FFMD well reproduced measured data which shows sharp mass-symmetric fission.
The calculated $Y(E_k)$ reasonably explains the measured data, including the largest yield at $E_k=\SI{230}{\MeV}$ and the tail distributed to lower TKE value.
In the calculation based on the five dimensional nuclear shape with the Metropolis method, FFMD and TKE distribution for $\nuclide{258}{Fm}$ were calculated in\ \cite{Albertsson2021-bd}.
Their results also show the pronounced peak structure in FFMD and clearly isolated two components in TKE distribution.

Our calculation of the 2D plot on the mass-TKE in\ \Cref{fig:masstke_massfke_fm258}(c) clearly shows the two fission modes,
i.e.\ symmetric fission with large TKE, centered at $E_k=\SI{230}{\MeV}$, and asymmetric fission with lower TKE, distributed around $(E_k,A_H)=(\SI{190}{\MeV},\SI{155}{\atomicmassunit})$.
The calculated high TKE values in symmetric fission distributes closer to the fission Q-value, obtained from the mass table\ \cite{Moller2016-cy} added by initial excitation energy $\SI{7.0}{MeV}$. 
The open circles represent the calculated average TKE for each fragment mass.

Figure\ \ref{fig:masstke_massfke_fm258}(d) shows the calculated distribution on the mass-FKE plane.
The most prominent difference of $\nuclide{258}{Fm}$ from $\nuclide{240}{Pu}$ is the increase of fragment energy in the symmetric fission.
For asymmetric fission, the trend is similar to $\nuclide{240}{Pu}$ and other actinide nuclides by showing that the average kinetic energy is nearly constant for light fragments, whereas the average FKE for heavy fragment decreases linearly with mass.

\begin{figure}[htb]
	\centering
	\includegraphics[width=1.0\linewidth]{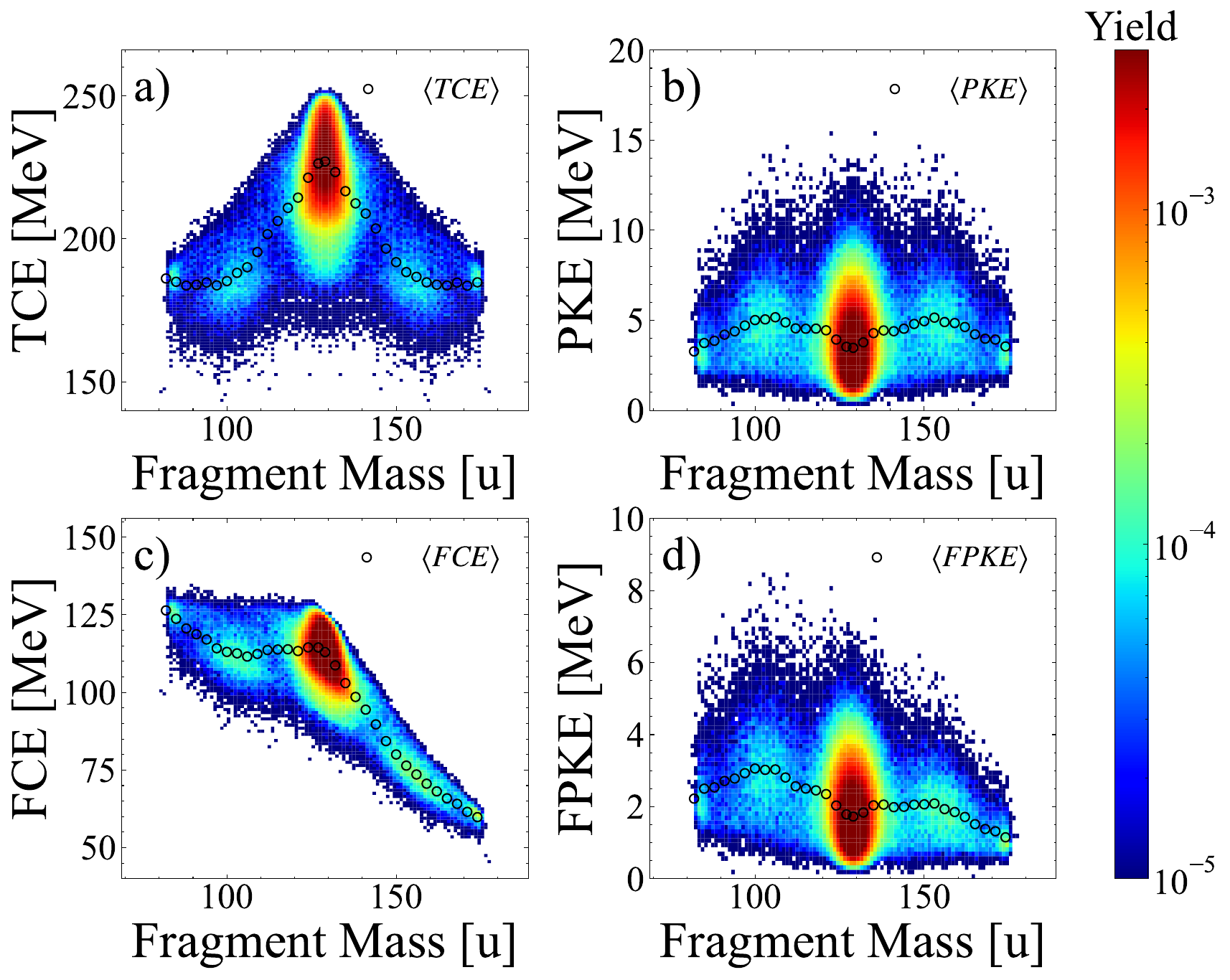}
	\caption{
		Same as \Cref{fig:masstce_masspke} but for $^{258}\mathrm{Fm}$ at $E^{*}=\SI{7.0}{\MeV}$.
	}\label{fig:masstce_masspke_fm258}
\end{figure}

Figure\ \ref{fig:masstce_masspke_fm258}(a) and (b) show mass-TCE and mass-PKE distributions for $\nuclide{258}{Fm}$.
The mass-TCE distribution is very similar to the mass-TKE distribution in \Cref{fig:masstke_massfke_fm258}(c),
indicating that most of the TKE is derived from the Coulomb energy.
Most of the mass-PKE events are distributed from 1 to \SI{11}{\MeV} for symmetric fission and from 3 to \SI{7}{\MeV} for asymmetric one.
The average PKE value fluctuates more largely with fragment mass than the fission of $\nuclide{240}{Pu}$ by showing local minimum at the symmetric fission.
Considering the \SI{4}{\MeV} of TKE in average over the fragment, it is estimated that PKE contributes about 2.2\% to TKE.
This is about half-value of the fission of $\nuclide{240}{Pu}$(4.2\%).

Figure\ \ref{fig:masstce_masspke_fm258} also shows the (c) mass-FCE and (d) mass-FPKE distributions.
For heavy fragments, FCE linearly decreases with mass similar to $\nuclide{240}{Pu}$.
For the average FCE of light fragments, it distributes around \SI{113}{\MeV} in the region of $90 < A_L < 120$.
More lighter fragments of $A_L < 90$ rather increases the FCE value.
Contrasted to $\nuclide{240}{Pu}$ shown in \Cref{fig:masstce_masspke}(c), the FCE of $\nuclide{258}{Fm}$ is locally enhanced in the symmetric fission.

In our calculation of $\nuclide{240}{Pu}$, pre-scission kinetic energy PKE has 4.2\% ($\SI{7}{\MeV}$) among TKE.
The $\nuclide{258}{Fm}$ gives 2.2\% ($\SI{4}{\MeV}$).
They are enough small compared to the fission Q-value ($\SI{220}{\MeV}$ for $\nuclide{240}{Pu}$ and $\SI{250}{\MeV}$ for $\nuclide{258}{Fm}$ at maximum values).
The rest of the energy is used in Coulomb energy, fragment deformation, and intrinsic excitation energy at the scission point.
Even having PKE predicted in our calculation, the system is very flexible to have any scission configuration.

\subsection{Evolution of pre-scission kinetic energy}

\begin{figure*}[ht]
	\centering
	\subfloat[$\nuclide{240}{Pu}$, asymmetric fission]{
		\includegraphics[height=0.47\textheight]{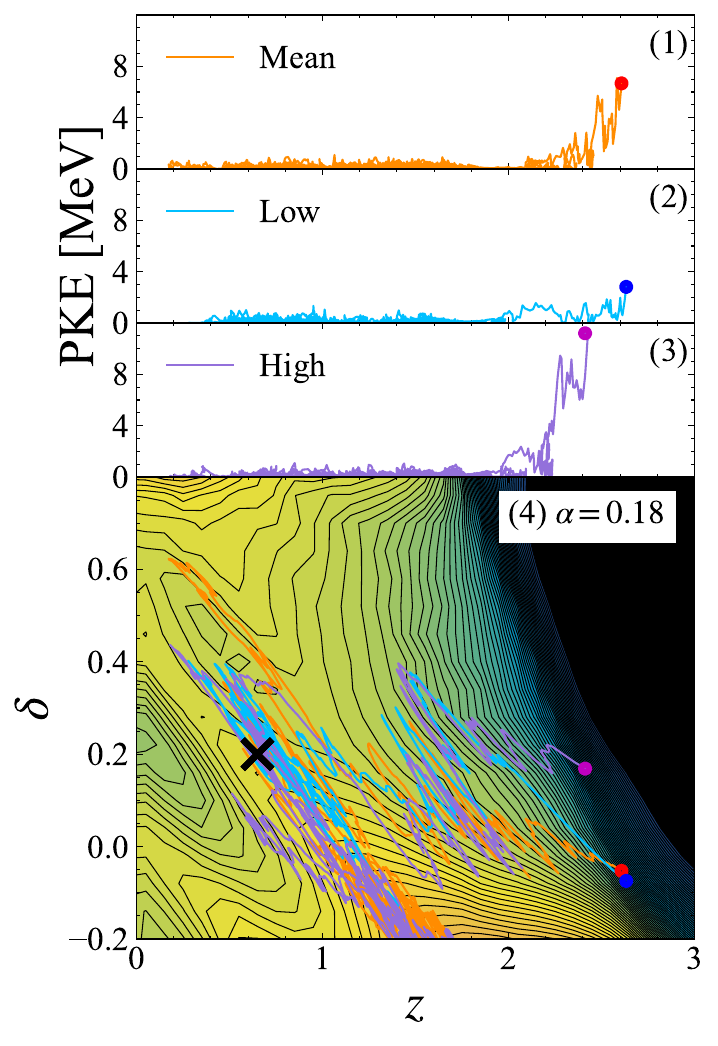}\label{fig:ek_zdep_pu240_sub}
	}
	\subfloat[$\nuclide{258}{Fm}$, symmetric fission]{
		\includegraphics[height=0.47\textheight]{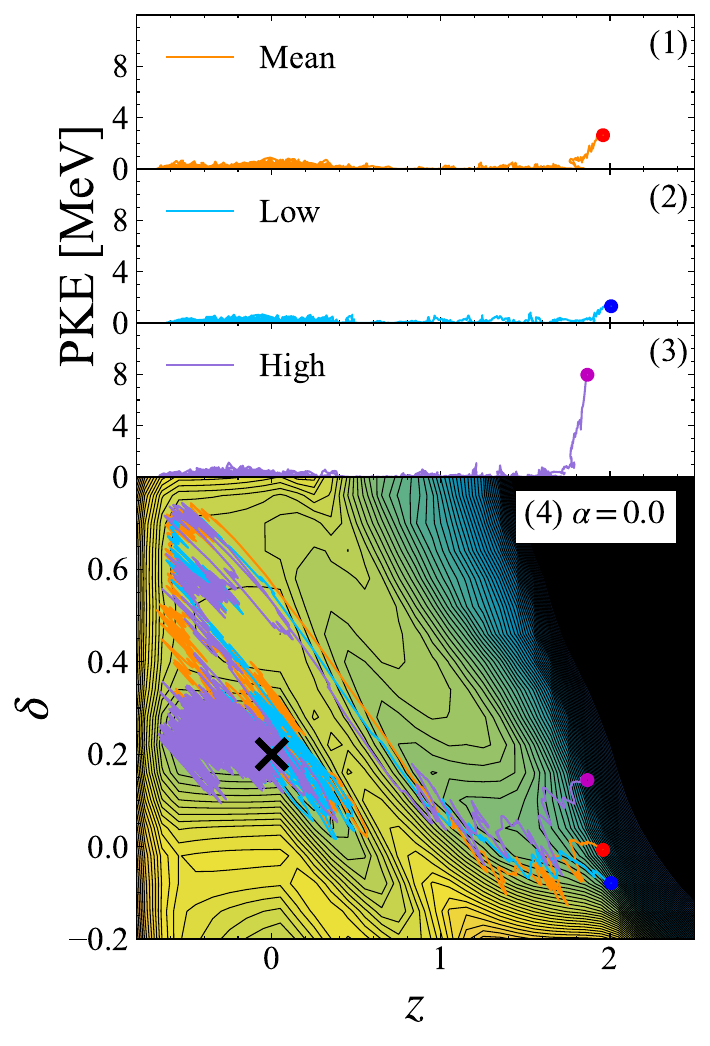}\label{fig:ek_zdep_fm258_sub}
	}
	\subfloat{
		\includegraphics[height=0.26\textheight]{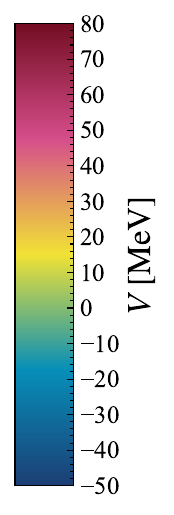}
	}
	\caption{
        \protect\subref{fig:ek_zdep_pu240_sub} Evolution of PKE is plotted as a function of $z$ when \nuclide{240}{Pu} splits mass asymmetry (three upper panels), where `High', `Mean', and `Low' show the line which has the high, middle, and low PKE value at the scission point.
        The lowest panel shows their trajectories on the $z-\delta$ plane $(\alpha=0.18)$.
        \protect\subref{fig:ek_zdep_fm258_sub} Same as \protect\subref{fig:ek_zdep_pu240_sub} but for \nuclide{258}{Fm}.
        Trajectories are chosen when mass-symmetric fission occurs.
        The potential energy surface is given for $\alpha=0$.
	}\label{fig:ek_zdep}
\end{figure*}

In order to see how pre-scission kinetic energy (PKE) is generated in the fission process,
we calculated the evolution of PKE with $z$ as shown in \Cref{fig:ek_zdep} for both system of (A) \nuclide{240}{Pu} and (B) \nuclide{258}{Fm} (upper three panels).
Three cases are selected, depending on the final PKE reached for each system.
For fission of \nuclide{240}{Pu} `Mean', `Low', and `High' are the results when the fragments have ($A_H$, TKE, PKE) values of (\SI{146.2}{\atomicmassunit}, \SI{174.4}{\MeV}, \SI{7.1}{\MeV}), (\SI{139.7}{\atomicmassunit}, \SI{172.0}{\MeV}, \SI{1.9}{\MeV}), and (\SI{136.2}{\atomicmassunit}, \SI{159.6}{\MeV}, \SI{11.2}{\MeV}), respectively.
For \nuclide{258}{Fm}, ($A_H$, TKE, PKE) = (\SI{133.4}{\atomicmassunit}, \SI{230.6}{\MeV}, \SI{2.5}{\MeV}), (\SI{129.5}{\atomicmassunit}, \SI{240.7}{\MeV}, \SI{1.6}{\MeV}), and (\SI{131.3}{\atomicmassunit}, \SI{215.1}{\MeV}, \SI{8.3}{\MeV}) are shown.
For \nuclide{240}{Pu}, the system has almost no PKE value until it reaches the $z = 2.0 - 2.2$.
\nuclide{258}{Fm} can have PKE value when the $z$-value of $1.7 – 1.9$ is reached.

\begin{figure}
	\centering
	\subfloat[$\nuclide{240}{Pu}$]{
		\includegraphics[width=0.2\textwidth]{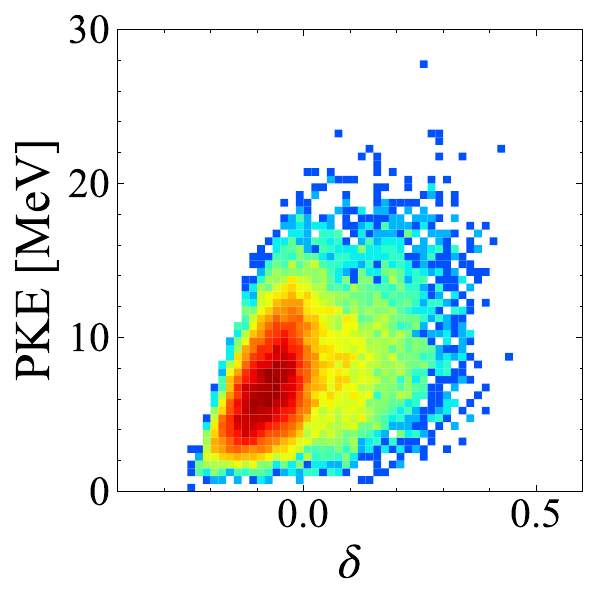}\label{fig:pke_delta_pu240}
	}
	\subfloat[$\nuclide{258}{Fm}$]{
		\includegraphics[width=0.2\textwidth]{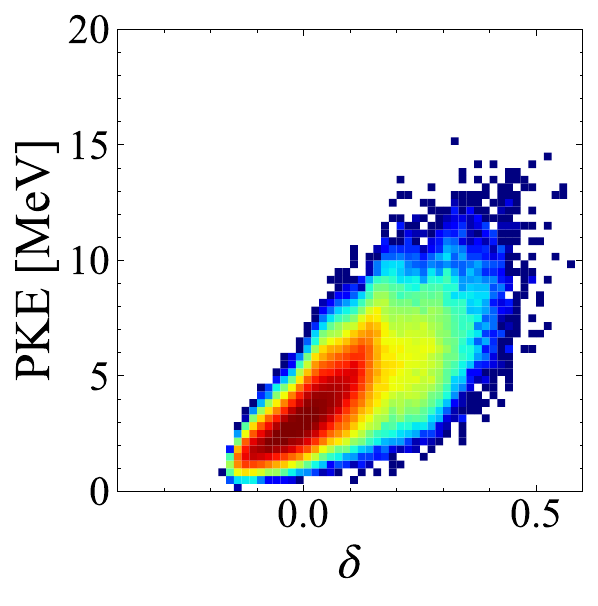}\label{fig:pke_delta_fm258}
	}
    \subfloat{\includegraphics[width=0.055\textwidth]{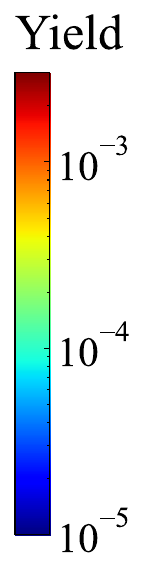}}
	\caption{
        Events recorded on the calculated PKE value and $\delta$ value at the scission point for fission of \protect\subref{fig:pke_delta_pu240} \nuclide{240}{Pu} $(E^* = \SI{6.53}{\MeV})$ and \protect\subref{fig:pke_delta_fm258} \nuclide{258}{Fm} $(E^* = \SI{7.0}{\MeV})$.
	}\label{fig:pke_delta}
\end{figure}

In the lowest panel of \Cref{fig:ek_zdep}, we show the potential energy landscape and trajectories corresponding to the selected three events.
The $z$-value that the system can start to accumulate when the PKE is positioned at the very low potential energy of \SI{-16.1}{\MeV} (\nuclide{240}{Pu}, `High'-trajectory) and \SI{-6.2}{\MeV} (\nuclide{258}{Fm}, `High'-trajectory).
The final PKE is close correlation with the movement of fragments in the vicinity of the scission point.
When the nuclear shape is developed to have large $\delta$ value, the fragments have a larger PKE.
For the case of low PKE, the trajectory tends to have a  smaller $\delta$ value as $z$ increases.
In \Cref{fig:pke_delta}, we show the relation between $\delta$ value and PKE obtained using all the registered trajectory events.
Here, $\delta$ value is chosen at the scission point.
Clear correlation between $\delta$ and PKE are noticed.
The result reveals that fragment tends to have a small pre-scission kinetic energy when the compact  scission configuration is reached.

\section{summary}

The kinetic energy of individual fragments calculated by the three-dimensional Langevin equation.
In addition to the static Coulomb energy at the scission point, we evaluated the pre-scission kinetic energy of fragments,  realized in our dynamical approach.
We studied thermal-neutron induced fission of $\nuclide{239}{Pu}$ and fission of $\nuclide{258}{Fm}$ from the low excited state.
For fission of $\nuclide{240}{Pu}$ the calculation reproduced the experimentally observed average kinetic energy for individual fission-fragment as a function of fragment mass, especially the nearly constant distribution for light fragments and linearly decreasing trend with heavy fragment mass is demonstrated.
Most of the TKE value originates from the Coulomb energy at the scission point, and the contribution from the pre-scisson kinetic energy is marginal.
We also note the calculation reproduced the experimental data of fission-fragment mass distributions.
For fission of \nuclide{258}{Fm} ($E^{*}=\SI{7.0}{MeV}$), calculation well reproduced the experimental data; sharp mass-symmetric FFMD and two mode fission found in TKE measurement.
The calculation shows that individual fragment kinetic energy follows nearly the same trend as low energy fission of \nuclide{240}{Pu} in asymmetric fission region.
For symmetric fission, it gives local enhancement due to strong Coulomb energy given by the compact scission configuration.
So far experimental data on the kinetic energy of individual fragments are not available. Such a measurement is highly interesting. 

\section*{Acknowledgements}
The Langevin calculations were performed using the cluster computer system (Kindai-VOSTOK)
which is supported by the Japan Society for the Promotion of Science (JSPS) KAKENHI Grant Number 20K04003.

\bibliography{fke}

\end{document}